\newif\ifarxiv
\arxivtrue

\ifarxiv
  \documentclass[conference]{IEEEtran}
\else
  \documentclass{KERauth}
\fi

\usepackage{url}
\usepackage{graphicx}

\ifarxiv
    \usepackage[nocompress]{cite}
\else
  \usepackage[round, authoryear,firstinits=true,urldate=long]{}
  \usepackage[backend=bibtex]{biblatex}
  \bibliography{ref.bib}
\fi

\begin{document}

\ifarxiv
  \bstctlcite{IEEEexample:BSTcontrol}
\fi

\title{The merits of using Ethereum MainNet as a Coordination Blockchain for Ethereum Private Sidechains}

\ifarxiv
  \author{
    \IEEEauthorblockN{Peter Robinson}
    \IEEEauthorblockA{Protocol Engineering Group and Systems (PegaSys), ConsenSys\\
    peter.robinson@consensys.net}
    \IEEEauthorblockA{School of Information Technology and Electrical Engineering, University of Queensland, Australia\\
    peter.robinson@uqconnect.edu.au}
  }
  \maketitle
  
  \thispagestyle{plain}
  \pagestyle{plain}
\fi

\begin{abstract}
\ifarxiv
A \textit{Coordination Blockchain} is a blockchain with the task of coordinating activities of multiple private blockchains. This paper discusses the pros and cons of using Ethereum MainNet, the public Ethereum blockchain, as a Coordination Blockchain. The requirements Ethereum MainNet needs to fulfil to perform this role are discussed within the context of Ethereum Private Sidechains, a  private blockchain technology which allows many blockchains to be operated in parallel, and allows atomic crosschain transactions to execute across blockchains. Ethereum MainNet is a permissionless network which aims to offer strong authenticity, integrity, and non-repudiation properties, that incentivises good behaviour using crypto economics. This paper demonstrates that Ethereum MainNet does deliver these properties. It then provides a comprehensive review of the features of Ethereum Private Sidechains, with a focus on the potential usage of Coordination Blockchains for these features. Finally, the merits of using Ethereum MainNet as a Coordination Blockchain are assessed. For Ethereum Private Sidechains, we found that Ethereum MainNet is best suited to storing long term static data that needs to be widely available, such as the Ethereum Registration Authority information. However, due to Ethereum MainNet's probabilistic finality, it is not well suited to information that needs to be available and acted upon immediately, such as the Sidechain Public Keys and Atomic Crosschain Transaction state information that need to be accessible prior to the first atomic crosschain transaction being issued on a sidechain. Although this paper examined the use of  Ethereum MainNet as a Coordination Blockchain within reference to Ethereum Private Sidechains, the discussions and observations of the typical tasks a Coordination blockchain may be expected to perform are applicable more widely to any multi-blockchain system.

\fi

\end{abstract}

\ifarxiv
\IEEEpeerreviewmaketitle
\fi

\section{Introduction}
This paper analyses the advantages and disadvantages of using Ethereum MainNet as a \textit{Coordination Blockchain}, by demonstrating the ways in which a Coordination Blockchain may be leveraged in a blockchain network that runs several parallel blockchains. We conduct an in-depth review of the features of Ethereum Private Sidechains, which is an example of such a blockchain system, to explore their potential usage of Coordination Blockchains as an exposition of using Coordination Blockchains more generally. The analysis builds on the Symposium on Distributed Ledger Technology paper \textit{Future of Blockchain} \cite{robinson2018d}, and other work on Ethereum Private Sidechains including: \textit{Requirements for Ethereum Private Sidechains} \cite{robinson2018a}, \textit{Ethereum Registration Authorities} \cite{robinson2018b}, \textit{Anonymous Pinning} \cite{robinson2019a}, and \textit{Atomic Crosschain Transactions} \cite{robinson2019b}. 

Ethereum MainNet is the largest public deployment of the Ethereum platform. It is a permissionless network, allowing any node to join the network. It is said to offer good authenticity, integrity, and non-repudiation properties, along with an economic system to discourage transaction spamming \cite{xu2017, griffith2018}. To date there has been no work that has analysed all of these assertions. This paper remedies this deficiency by carefully analysing whether these properties are successfully delivered. 

Sidechains are blockchains that rely on a separate blockchain, a Coordination Blockchain, for their overall utility. This could be to enhance security by pinning the state of the sidechain to the Coordination Blockchain \cite{robinson2019a}, for addressing information \cite{robinson2018b}, or for storing data that is used across all sidechains. We analyse the appropriateness of using Ethereum MainNet as a Coordination Blockchain for the various features of sidechains, using as a reference Ethereum Private Sidechains.

This paper is organised as follows: the \textit{Background} section briefly introduces Ethereum MainNet, the platform that forms the basis for this paper. We describe the concept of private blockchains and the enterprise version of Ethereum, and introduce the concept of block `finality'. Next cryptanalysis of message digest and asymmetric algorithms is reviewed given classical and quantum cryptanalytical techniques. The \textit{Ethereum MainNet Features} section analyses whether Ethereum MainNet delivers authenticity, integrity, non-repudiation, and crypto-economic anti-spam properties. The \textit{Ethereum Private Sidechains} section describes the features of Ethereum Private Sidechains and their usage of Coordination Blockchains. The \textit{Pros and Cons of using Ethereum MainNet as a Coordination Blockchain} section analyses the advantages and disadvantages of using Ethereum MainNet as the Coordination Blockchain for each of the Ethereum Private Sidechain features.

\section{Background}
\subsection{Ethereum}
\subsubsection{Ethereum MainNet}
\label{ref:ethereum}
Ethereum \cite{wood2016a} is a blockchain platform that allows users to upload and execute computer programs known as Smart Contracts. Ethereum Smart Contracts can be written in a variety of Turing Complete languages, the most popular being Solidity \cite{solidity}. Source code is compiled into a bytecode representation. The bytecode can then be deployed using a contract creation transaction. Contracts have a special \textit{constructor} function that only runs when the contract creation transaction is being processed. This function is used to initialize memory and call other contract code. Miners execute the bytecode inside the Ethereum Virtual Machine (EVM). At present, each miner must execute all transactions for all contracts and hold the current value of all the memory associated with all of the contracts. The Ethereum community is actively working on methodologies to scale the Ethereum network by sharding the blockchain \cite{ethereum-sharding}.

Ethereum transactions update the state of the distributed ledger but do not return values. They fall into three categories: Ether transfer, contract creation, and calling a function on a contract. Ether transfer transactions move Ether from the user's account to another account. Contract creation transactions put code into the distributed ledger and call the constructor of the contract code, setting the contract data's initial state. Function call transactions call a function on a contract and result in updated state. Contract creation and function call transactions also allow Ether to be transferred. All types of transactions must be signed by a private key corresponding to an account and include a nonce value that prevents replay attacks. In addition to Ethereum transactions, ``View" function calls can be executed on the Smart Contract code. These View function calls return a value and do not update the state of the Smart Contract.

Executing code and accessing resources, such as memory, costs certain amounts of ``Gas". The ``Gas Cost'' of executing code is closely tied to the real world cost of executing each type of instruction. Miners preferentially mine transactions that are prepared to pay a higher ``Gas Price''. Accounts instigating transactions specify the ``Gas Price" they are prepared to pay for their transaction and specify the maximum amount of gas a transaction can use known as ``Start Gas". This commits an account holder to paying up to a certain amount of Ether for the transaction. Any unused gas is returned to the account holder at the end of the transaction. Transactions that run out of gas prior to completion are aborted, with all of the gas being expended.

In the Ethereum public network, ``MainNet", all contract code and data are readable by any user of any node that connects to the network. Smart Contracts on Ethereum MainNet can only perform permissioning in contract code, limiting which accounts can update the state of a contract. However, there is no mechanism to limit which users can read contract code and data.

\subsubsection{Private Blockchains and Enterprise Ethereum}
Private blockchains are blockchain networks that are established between nodes operated by enterprises \cite{robinson2018a}. Only permissioned nodes belonging to participating enterprises are allowed to join the private blockchain's peer-to-peer network and only permissioned accounts belonging to participating enterprises are allowed to submit transactions to the nodes. These blockchains provide the privacy and permissioning required by enterprises \cite{enteth20}. 

The need for security and permissioning features over and above what is available in standard Ethereum \cite{enteth20} has led to a range of platforms being developed. J.P. Morgan developed Quorum \cite{quorum-source}, a fork of the Golang Ethereum implementation called Geth  \cite{geth-github}. ConsenSys's Protocol Engineering Group, PegaSys created Pantheon \cite{pantheon-github}, an Ethereum MainNet compatible client that aims to meet the permissioning and privacy requirements of the Enterprise Ethereum Client Specification \cite{enteth20}. Hyperledger Fabric \cite{androulaki2018} is a distributed ledger platform originally created by IBM and now hosted by The Linux Foundation. Similar to Quorum and Pantheon, the platform offers privacy and permissioning features. Whereas Quorum offers Ethereum based private transactions, Pantheon offers private smart contracts that are private to a set of participants. Hyperledger Fabric offers the ability to host one or more smart contracts on a private blockchain called a ``channel". Hyperledger Fabric allows multiple channels to be operated on the one network, thus allowing for multiple sets of private contracts between different sets of participants to operate on the one network. An analysis of the merits of Hyperledger Fabric and Quorum has been analysed elsewhere (see \textit{Requirements for Ethereum Private Sidechains} \cite{robinson2018a}).

\subsubsection{Finality}
\label{section:finality}
A block is deemed \textit{final} when it can no longer be changed. All transactions contained within a finalised block are also deemed final.

Ethereum transactions are included in blocks. An Ethereum MainNet miner that solves the Proof of Work cryptographic puzzle can add a block to the end of the blockchain. If two or more miners solve the puzzle simultaneously, then two or more chains are created with common ancestors, and this is known as a \textit{fork} \cite{DeAngelis2018}. In Bitcoin the longest chain of blocks is deemed to be the valid blockchain
\cite{nakamoto2008,Courtois2014}. In Ethereum, the fork choice is solved by means of a modified Greediest Heaviest Observed Subtree (GHOST) protocol \cite{DeAngelis2018} that takes into account the mining power in creating blocks that have links to the main chain, but have become stale \cite{Gencer2018}. These blocks are commonly referred to as \textit{uncle} blocks. The \textit{weight} of a block relates to the number of previous blocks in the chain and uncle blocks. The heaviest chain of blocks is deemed to be the valid blockchain. If an Ethereum MainNet miner becomes aware of a heavier chain than it knew about, it should then only attempt to add blocks to the new chain. Blocks on the old heaviest chain that are not in common with the new longest chain are deemed \textit{reordered}. If none of the transactions in a reordered block have been included in the blocks of the new longest chain, then the block can be included as an \textit{uncle} block. Otherwise, the transactions that are not included in the reordered chain need to be included in a new block. There is no certainty that these transactions will be included in a new block, or that transactions in a proposed uncle block will be included in the blockchain. 

As more blocks are added to the end of Ethereum MainNet's blockchain, the probability of a miner finding a longer blockchain and reordering the blockchain is reduced \cite{DeAngelis2018}. This is because a miner would need to repeatedly solve the Proof of Work cryptographic puzzle for each block faster than all other miners. As the probability of a block being reordered is reduced, the probability of the transactions included in a block being \textit{final} increases. Hence, Ethereum MainNet is said to have, \textit{probabilistic finality} \cite{Courtois2014}.

Consensus algorithms such as Istanbul Fault Byzantine Tolerant (IBFT) \cite{ibft} and Istanbul Fault Byzantine Tolerant version 2 (IBFT2) \cite{ibft2} used in consortium blockchains give \textit{instant} finality, where once a transaction has been included in a block minted by a validator, it can no longer be changed.

\subsubsection{Pinning}
\begin{figure}
 \includegraphics[width=\linewidth]{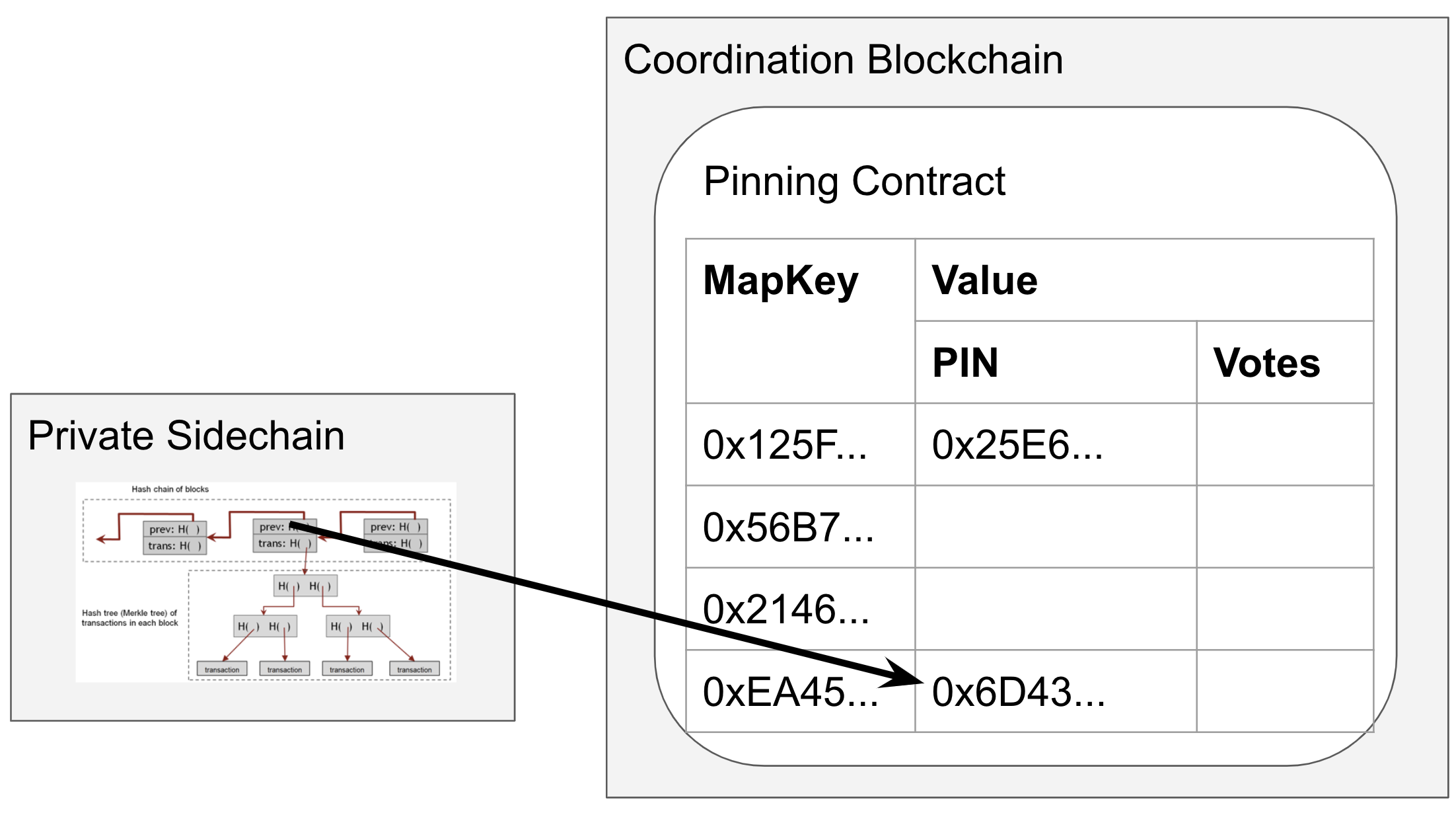}
 \caption{Pinning}
  \label{fig:pinning}
\end{figure}

The state of a blockchain or sidechain can be represented by the Block Hash of a block. The Block Hash of a final block could be submitted to a contract on a Coordination Blockchain at regular intervals  \cite{robinson2019a}, as shown in Figure \ref{fig:pinning}. This process is know as \textit{pinning}. Regularly pinning sidechain state helps to protect minority sidechain participants from state reversion due to collusion by the majority of sidechain participants \cite{robinson2019a}. 

\subsection{Cryptanalysis}
This section provides background material on cryptanalysis that is needed to understand the analysis of the security properties of Ethereum MainNet. 

\subsubsection{Message Digest Algorithm Cryptanalysis}
Message digest algorithms have three main security properties: Preimage Resistance, Second Preimage Resistance, and Collision Resistance. Message digest algorithms are commonly called Cryptographic Hash algorithms, or simply Hash algorithms. Given a Hash algorithm \textit{h}, the three security properties can be stated as:
\begin{itemize}  
\item \textit{Preimage Resistance}: Given $y$, it is difficult to determine $x$ such that $y = h(x)$.
\item \textit{Second Preimage Resistance}: Given $y$ and $x_1$, it is difficult to determine $x_2$ such that $y = h(x_1) = h(x_2)$ and $x_1 \neq x_2$.
\item \textit{Collision Resistance}: It is difficult to determine $x_1$ and $x_2$ such that $h(x_1) = h(x_2)$ and $x_1 \neq x_2$.
\end{itemize}

\subsubsection{Classical Computing Cryptanalysis}
\label{section-classical}
Gordon Moore, co-founder of Intel, stated in his ''Moore'€™s Law'' that the number of transistors on an integrated circuit doubles approximately every two years \cite{moore1995}. With the increased number of transistors has come a decrease in transistor size, which has resulted in decreased power consumption per transistor. This has resulted in an increase in computation power, while keeping the power consumption relative static over a fifty year period. This rate of increase of computation power and decrease of transistor size though slowing, is still continuing \cite{cunningham2016}. Additionally, new alternative approaches are being developed to deliver increased computational power \cite{simonite2016}.

Classical computational power can be used to break algorithms such as message digest algorithms by trying all possible combinations using a ``Brute Force" attack. Complexity theory predicts how many attempts are likely to be needed to break an algorithm. For message digest algorithms, using classical computing power, the complexity of breaking an algorithm's Preimage Resistance or Second Preimage Resistance property is $O(N)$, where $N$ is the number of combinations of the digest output, whereas the complexity of breaking an algorithms Collision Resistance is $O(\sqrt{N})$. 

The USA's National Institute of Standards and Technology (NIST) defines \textit{Security Strength} \cite{barker2016a} as, ``A number associated with the amount of work (that is, the number of operations) that is required to break a cryptographic algorithm or system.'' Security Strength and complexity are related. The Security Strength of a message digest algorithm's Preimage and Second Preimage Resistance properties is $\log_2{N}$ and the Collision Resistance Security Strength is $\log_2{\sqrt N}$. Recall that $\log_2{N}$ corresponds to the message digest output length in bits. As such, the algorithm \texttt{SHA-256}'s Preimage and Second Preimage Security Strength is 256-bits and its Collision Resistance Security Strength is 128 bits, assuming classical computers \cite{barker2016a}.

In some instances, a message digest output is truncated. For example in Ethereum, \texttt{Keccak-256} is used to generate account numbers with the output truncated from 256-bits to 160-bits. In this usage, the analysis of Security Strength remains unchanged: the complexity and hence Security Strength relates to the number of possible values of the digest output. If a message digest output is truncated then the Security Strength of the overall algorithm is proportionally reduced.

NIST defines algorithms with Security Strengths of 80, 112, 128, 192, and 256 bits \cite{barker2015a}. NIST have mandated the phasing out of 80-bit Security Strength algorithms in 2010 and, based on Moore'€™s Law, had indicated the phasing out of 112-bit Security Strength algorithms by 2030.

\subsubsection{Quantum Computing Cryptanalysis}
\label{section-quantum}
Quantum computers are expected to allow all currently used popular asymmetric cryptographic algorithms to be defeated and are expected to reduce the Security Strength of message digest and symmetric cipher cryptographic algorithms \cite{nist2016a}. Aggarwal et al. \cite{ledger127} estimate that ECC 256-bit schemes will be able to be compromised with a Quantum computer using the Shor algorithm \cite{shor1994} in less than ten minutes sometime between 2027 and 2040. 

Grover's algorithm \cite{grov1996} provides a speedup for database search style algorithms, such as searching for a message digest preimage or second preimage. Using Grover's algorithm the complexity of message digest algorithm's Preimage or Second Preimage Resistance properties are reduced from $O(N)$ to $O(\sqrt{N})$. This means that the Security Strength assuming a sufficiently powerful quantum computer is half that when compared to the Security Strength due to classical computing power.

Brassard and Tapp \cite{Brassard1997} claimed to have developed an algorithm for use with quantum computers that reduces the complexity of finding message digest collisions to $O(\sqrt[3]{N})$. Bernstein \cite{Bernstein2009} has refuted this claim, stating that there is no real advantage provided by Brassard and Tapp's algorithm given the cost - performance analysis over classical computing power. However, Aaronson and Shi \cite{Aaronson2004} have determined a tight lower bound for the complexity of the collision problem as $O(\sqrt[3]{N})$. As such, despite Bernstein's refutation of Brassard and Tapp's algorithm, it can be conjectured that another algorithm may be found that meets the theoretical bound, that has a better cost - performance metric.

Despite the reduced Security Strength offered by message digest algorithms, assuming a quantum computer, they are unlikely to be a point of weakness in the near term. Developing a complex quantum computer that can defeat message digest algorithms is expected to be significantly more complex than developing one to defeat \texttt{ECC} 256-bit \cite{mosca2015}. As such, it is likely that a quantum computer that can be used to attack message digest algorithms will not be available until at least the 2030s.

\subsubsection{Algorithmic Weaknesses}
Researchers search for weaknesses in algorithms. These weaknesses when found can reduce the effective Security Strength offered by the algorithm. For example various weakness have been found in the \texttt{MD-5} message digest algorithm \cite{Wang2005} \cite{Sasaki2009}. It is impossible to predict if a weakness in an algorithm such as \texttt{Keccak-256} will be found, and the degree to which the algorithm would be weakened with such a compromise. Algorithmic weaknesses will not be considered in the analysis of Ethereum MainNet given the uncertainty as to whether such weakness will be found, when they will be found, and the impact such weaknesses might have.

\section{Ethereum MainNet Features}
This section discusses in detail the features of Ethereum MainNet that are important to its usage as a Coordination Blockchain.

\subsection{Authentication}
The International Telecommunications Union (ITU) define \textit{authentication} in X.805 \cite{x805} as:
\begin{quote}
...serves to confirm the identities of communicating entities. Authentication ensures the validity of the claimed identities of the entities participating in
communication (e.g., person, device, service or application) and provides assurance that an entity is not attempting a masquerade or unauthorized replay of a previous communication. 
\end{quote}
In the context of Ethereum, this means ensuring Ethereum transactions are directly attributable to participants who operate Ethereum Accounts.

Ethereum transactions are signed using the private key belonging to a participant \cite{wood2016a}. The public key associated with the private key can be derived from the transaction signature of any transaction signed by the private key. The account number is the twenty-byte truncated \texttt{Keccak-256} message digest of the public key. 

In Ethereum, each transaction includes a \textit{nonce} \cite{wood2016a}. The initial nonce value for each account is zero. The nonce is incremented for each successfully mined transaction. Miners reject transactions with out of order or repeated nonces. Doing this protects Ethereum from transaction replay attacks.

The nonce value is represented as a 64-bit signed number in Geth \cite{geth-github} and Pantheon \cite{pantheon-github}. Adding one to the maximum representable number would result in the largest negative number. If this situation was not guarded against in the code, it would lead to unexpected results, and possibly an authentication failure. However, 63-bits is large enough such that even if a single account issued every transaction on Ethereum MainNet, and could craft sufficiently small transactions and could have the gas limit increased such that they could execute 1000 transactions per second, the nonce value would not wrap around for 584 million years.

Ethereum private keys are 256-bits long. The signature algorithm \texttt{ECDSA / Keccak-256} using the \texttt{secp256k1} curve is used for signing transactions. The \texttt{secp256k1} curve has been analysed and found to not have any weaknesses \cite{ecdsacurve}. This signature algorithm provides 128-bits of Security Strength \cite{barker2015a} assuming Classical Cryptanalysis. The conversion of the public key to an account number using a twenty-byte truncated \texttt{Keccak-256} message digest offers 160-bits of Security Strength assuming Classical Cryptanalysis, as an attacker would need to exploit the Second Preimage Resistance property of the message digest function to determine another public key which could hash to the same value as the authentic public key. As such, overall the Ethereum signing mechanism provides 128-bits of Security Strength assuming Classical Cryptanalysis.

NIST has issued guidance that usage of algorithms offering 112-bit Security Strength assuming Classical Cryptanalysis should be phased out by 2030 \cite{barker2016a}. This means that Ethereum's transaction signing technique should be secure well beyond 2030, assuming Classical Cryptanalysis, given its 128-bit Security Strength.

If an attacker had access to a sufficiently powerful Quantum Computer, they could determine private keys associated with the public keys. The attacker could observe transactions that have been submitted and determine the public keys associated with each transaction using the standard \texttt{ecrecover} technique \cite{wood2016a}. Once an attacker had access to a private key, they could issue arbitrary transactions using that private key. Aggarwal's \cite{ledger127} analysis indicates that the authenticity of transactions may be able to be compromised in this way some time after 2027. 

The Ethereum community have recognised the threat that Quantum Cryptanalysis poses to Ethereum transaction signing. There are plans to roll-out ``Account Security Abstraction" changes that will authenticate transactions programmatically using user supplied code \cite{buterin2015a} \cite{buterin2016c}\cite{eip86b}. This would allow for users to choose to use Quantum Cryptanalysis resistant algorithms.

In summary, the existing transaction authentication techniques are likely to be secure until at least 2027. Prior to 2027, Ethereum is likely to be upgraded to mitigate the threat of quantum computers, thus ensuring the authenticity of transactions into the future.

\subsection{Integrity}
ITU defines \textit{data integrity} \cite{x805} as:
\begin{quote}
... ensures the correctness or accuracy of data. The data is protected against unauthorized modification, deletion, creation, and replication and provides an indication of these unauthorized activities.
\end{quote}
In the context of Ethereum, this means ensuring that authenticated transactions and data in the distributed ledger are stored such that they can not be modified.

Ethereum transactions are combined into blocks using Merkle Patricia trees \cite{wood2016a}. Similarly, data in the distributed ledger is protected using Merkle Patricia trees. Compromising values in the Merkle Particia trees would require breaking the Second Preimage Resistance property of \texttt{Keccak-256}. This is unlikely to occur in foreseeable future using either Quantum or Classical Cryptanalysis techniques. However, there is always the possibility that a weakness in \texttt{Keccak-256} will be found.

\subsection{Non-Repudiation}
\label{nonrep}
ITU defines \textit{non-repudiation} \cite{x805} as:
\begin{quote}
...provides means for preventing an individual or entity from denying having performed a particular action related to data by making available proof of various network-related actions (such as proof of obligation, intent, or commitment; proof of data origin, proof of ownership, proof of resource use). It ensures the availability of evidence that can be presented to a third party and used to prove that some kind of event or action has taken place. 
\end{quote}
In the context of Ethereum, this means ensuring that authenticated transactions are stored such that they can not be revoked.

Ethereum blocks are linked together using \texttt{Keccak-256} message digests. Compromising this linkage would require breaking the Preimage Resistance property of \texttt{Keccak-256}, which is unlikely to occur in foreseeable future. 

As discussed in Section \ref{section:finality}, {Finality}, Ethereum MainNet has \textit{probabilistic finality}. When blocks are added to the blockchain after a block containing a transaction, the probability of a miner proposing a heavier chain that does not include the block decreases. The number of blocks added after a block is known as the number of \textit{block confirmations}. Nakamoto \cite{nakamoto2008} showed the probability of a Bitcoin block being removed after six blocks, assuming an attacker has 10\% of the mining power was \texttt{0.00024}. A greater number of \textit{block confirmations} should be observed if an attacker were assumed to have a greater percentage of the total mining power available to them, or if the user wished to have greater certainty that the block was not going to be removed.

In 2016, Gervais \cite{Gervais:2016} determined that 37 Ethereum MainNet block confirmations were needed to offer the same level of security as six Bitcoin block confirmations, assuming Ethereum was being attacked with 30\% of mining power. Since 2016, the mining power devoted to Ethereum has increased considerably such that a 30\% attack now seems inconceivable. Major miners are unlikely to attack their own network as this would risk devaluing the cryptocurrency they are mining \cite{bitcoinmag2019, niu2019}. The maximum hash power which can be rented in a straightforward way is 5\% \cite{renthashpower}. Purchasing hardware to generate 30\% hash power (174TH/s \cite{ethstats}) would cost in excess of US\$400 million \cite{gpucost}.

Scaling the results of Gervais's work \cite{Gervais:2016} based on the changed mining rewards of Bitcoin and Ethereum, the changed valuations, and allowing for a 10\% mining power attack, indicates that eight Ethereum block confirmations corresponds to six Bitcoin confirmations. Using a different methodology, Buterin \cite{confirmations} determined that six to twelve confirmations where required to deem a transaction final, depending on the level of risk a user was prepared to assume. 

Based on a fourteen second target block time and assuming twelve confirmations, a block on Ethereum MainNet could be deemed final in approximately three minutes. The finality time is not a precise number as the block time is randomly distributed with an average of fourteen seconds. 

When Ethereum MainNet client vendors and miners agree to changes in the Ethereum protocol, the system is updated via changes known as \textit{Hard Forks}. A Hard Fork requires all Ethereum MainNet client vendors to release updated software which will activate new functionality at a certain Ethereum MainNet block number. For the Spurious Dragon Hard Fork in November 2016 \cite{hardfork2016} the changes were implemented slightly differently. This resulted in the Ethereum MainNet blockchain forking for some hours \cite{hardforkupdate}. The fork is resolved once the vendors software has been corrected. However, it is possible that a transaction which was part of a block accepted into the fork which was discarded was reverted and not resubmitted to the blockchain. This type of forking and state reversion due to mismatched feature implementation is much less likely to occur now and in the future than it did in 2016 as Ethereum MainNet clients undergo significantly more review and testing than they did in 2016 \cite{ethtesting1, ethtesting2}.

If an attacker could dedicate 51\% of the total mining power to attacking the network, they would be able to mount a \textit{51\% Attack} \cite{fifty-one-percent-attack}. This would allow the attacker to rewrite the history of the blockchain. The three largest Ethereum MainNet miners could collude to mount such as attack. However, these miners are disincentivized to do such an attack as this would adversely affect confidence in Ethereum MainNet. This would lead to a dramatic drop in the value of Ether \cite{bitcoinmag2019, niu2019}, substantially decreasing the value of their Ether and their Ethereum infrastructure investments.

Though the Ethereum MainNet system typically can not be modified, after a re-entrancy bug was exploited in the DAO attack \cite{atzei2017}, the system was modified to reverse the results of the attack. Doing this caused some to question trust in blockchain systems and Ethereum MainNet in particular \cite{spode2017}. However, this type of \textit{irregular state change} \cite{khoo2016} to reverse the results of such an attack appear unlikely to occur again in Ethereum MainNet. Despite a bug in the Parity Wallet contract that resulted in hundreds of millions of dollars of funds becoming inaccessible, proposals to alter history to restore the funds have been refused \cite{johnson2018}\cite{eip999}.

\subsection{Crypto Economic Anti-Spam}
As described in Section \ref{ref:ethereum}, each transaction on Ethereum MainNet costs Gas to execute, which participants pay for with Ether. Ethereum MainNet currently aims to produce new blocks each 14 seconds with eight million Gas available for each block \cite{ethstats}. Each transaction has as a minimum cost, the transaction fee, that is currently \texttt{21,000} Gas. Simple balance transfers between accounts just cost the transaction fee, whereas complex function calls can cost more than a million gas. As the block gas limit is eight million, it means that no transaction can use more than eight million gas. This translates to Ethereum MainNet supporting between four transactions per minute and twenty-seven transactions per second. A typical simple transaction, \textit{adding a Pin to a pinning contract}, costs \texttt{64972} Gas \cite{robinson2019a}. Given the eight million Gas limit, \texttt{8.8} of these transactions could execute per second.

Participants are disincentivized from flooding the network with transactions as each transaction has an economic cost. The cost of Gas depends on the block utilisation \cite{ethgasstation}. Historically, the Gas price has spiked high when block utilisation has been high \cite{crypto-kitties}. If many entities attempted to issue \textit{adding a Pin to a pinning contract} transactions regularly, such that the block utilisation was high, then the cost of issuing the \textit{adding a Pin to a pinning contract} transactions would increase. This would incentivise the entities to find alternatives, such as reducing the frequency of submitting the transactions.

\subsection{Summary}
Based on the analysis in this section, it can be said that Ethereum MainNet contains transactions for which the authenticity and integrity is certain. Once twelve blocks have been appended to the block containing a transaction, the probability of the blockchain being reorganised such that the transaction is reverted is small. As such, Ethereum MainNet offers strong non-repudiation properties. Ethereum's Gas mechanism operates as an effective anti-spam tool.

\section{Ethereum Private Sidechains}
Ethereum Private Sidechains are Ephemeral, On-demand, Permissioned, Private, Confidential, blockchains that allow for Atomic Crosschain Transactions. They are Ephemeral in that they are created, they operate, and then they can be archived when they are no longer needed. Their On-demand nature allows them to be created when needed between parties that have no prior relationship. Permissioning ensures that only authorised nodes are able to join a sidechain. Their design is such that to the greatest extent possible, their membership and their transactions are kept Private. Confidentiality is ensured by encrypting the sidechain data when being communicated between nodes and stored on nodes. Atomic Crosschain Transactions enable transactions that update state across sidechains atomically.

Ethereum Private Sidechains have been described in terms of their requirements \cite{robinson2018a}, and aspects of their technology \cite{robinson2018b,robinson2019a,robinson2019b}. This paper is the first to present this technology holistically. Additionally, this paper introduces the idea of pinning the final state of a sidechain prior to archiving, thus allowing the sidechain to be reinstated if needed, and introduces the idea of using Ethereum MainNet gas pricing as a mechanism for rate control of Atomic Crosschain Transactions.

\subsection{Ephemeral}
\label{ref:ephemeral}
Ethereum Private Sidechains are \textit{Ephemeral}: they are created, they are used for a period, and then archived when they are no longer required. This limited lifespan matches many real world requirements, such as \textit{Letters of Credit} and other business deals, which have a limited lifespan. The ability to archive the blockchain data in a sidechain is in contrast to existing blockchain technologies that are designed to be operational indefinitely.

The life span of a sidechain could vary widely. For usages in which sidechains are used to deploy a contract and automatically negotiate a deal, it might only be needed for some minutes, hours or days. Other usages, such as an \textit{Oracle}, require a long or indefinite lifespan. Indefinite lifespans can be accommodated by never archiving the sidechain. 

While a sidechain is operational, the sidechain could be pinned to a Coordination Blockchain at regular intervals  \cite{robinson2019a}. Regularly pinning sidechain state helps to protect minority sidechain participants from state reversion due to collusion by the majority of sidechain participants \cite{robinson2019a}. 

A key aspect of Ephemeral sidechains is the requirement to be able to restart the sidechain after archiving. This can be achieved by pinning the last block of the sidechain to a Coordination Blockchain. Now that the Block Hash of the last block has been securely stored in the Coordination Blockchain, the state of the sidechain can then be stored offline. To restart the sidechain, the stored data is compared against the final Block Hash to confirm the correct state is being used to restart the sidechain.

\subsection{On-demand Between Parties with No Prior Relationship}
\label{ref:ondemand}
Ethereum Private Sidechains need to be able to be deployed between parties that have no prior relationship. That is, the parties need to be able to establish a sidechain without knowing each others' node IP addresses, cryptographic keys, or other information required to set-up a secure connection. Establishing sidechains in this dynamic way is in contrast to existing permissioned blockchains that are largely static systems that require complex set-up. For example, set-up of a Quorum \cite{quorum-source} network requires enode addresses (IP addresses and Ethereum account numbers) for each node to be shared out of band with all other nodes.  Adding new nodes to the network requires this sharing and manual intervention on each node.

The on-demand sidechain establishment is analogous to a user of a web browser establishing a secure connection with a web server by simply entering in a URL such as \url{https://example.com/}. The user does not know the IP address of the computer corresponding to \textit{example.com} or the public key that can be used to verify the communications emanating from \textit{example.com}. However, using the domain name, some initial trust, and the Domain Name Service (DNS) and Transport Layer Security (TLS) protocols, they are able to establish a secure connection. Similarly, Ethereum Private Sidechains need to be able to establish a secure sidechain using just domain names.

The Ethereum Registration Authorities system is a set of smart contracts that can be used to provide discoverable information to enable establishment of sidechains between organisations with no prior relationship \cite{robinson2018b}. A Coordination Blockchain could be used to locate the information using domain names that can be grouped according to different trust levels and different trust relationships. Moreover, a Coordination Blockchain that provides organisations with a secure, decentralized, censorship-resistant mechanism for storing information that can be located using domain names and grouped according to different trust levels and different trust relationships would overcome the limitations of previous technologies that did not provide the security and censorship resistance properties that users of blockchain technologies expect.

\subsection{Permissioned}
Ethereum Private Sidechains need to be operated by authorised nodes using authorised Ethereum accounts. These requirements match those of the Enterprise Ethereum Client Specification \cite{enteth20}. The implementation of these requirements do not use a Coordination Blockchain.

\subsection{Private}
Ethereum Private Sidechains should, to the greatest extent possible, keep their membership private from other sidechains they interact with and from any Coordination Blockchains they use to facilitate their actions.

\subsection{Confidential}
Ethereum Private Sidechains should encrypt their blockchain and state data such that the transaction information is kept confidential, both when it is communicated between nodes on a sidechain and when it is stored in a node's local data store. The implementation of this feature does not use a Coordination Blockchain.

\subsection{Atomic Crosschain Transactions}
\label{atomic}
Ethereum Private Sidechains technology needs to enable transactions that update state across sidechains atomically \cite{robinson2019b}. That is, if an Atomic Crosschain Transaction is across sidechains A, B, and C, then the state updates related to the transaction are either applied on all sidechains or ignored on all sidechains. A Coordination Blockchain holds a Crosschain Coordination Contract. This contract is used to indicate that an Atomic Crosschain Transaction has commenced, has been committed, or should be ignored. The contract acts as a common time-out reference for all sidechains and helps prevent denial of service attacks. The data in the Crosschain Coordination Contract needs to be available until the last sidechain using it has been archived. 

The Atomic Crosschain Transaction system uses threshold signatures to prove values across sidechains. The public key that corresponds to the private key shares held by each of the sidechain validators is known as a Sidechain Public Key. This key needs to be available to all sidechains that need to verify values coming from a sidechain. As such, this value should be stored on a Coordination Blockchain. The Sidechain Public Key needs to be re-generated and uploaded to the Coordination Blockchain each time a validator is added or removed from the sidechain. Assuming that sidechain membership is largely static, this regeneration and upload is likely to be a rare event.

\section{Pros and Cons of using Ethereum MainNet as a Coordination Blockchain}
The subsections below analyse the advantages and disadvantages of using Ethereum MainNet as the Coordination Blockchain for the operations of an Ethereum Private Sidechain. The findings of the subsections are summarised in Table \ref{table:merits}.

\begin{table*}[t]
  \centering
  \begin{tabular}{| l | l | l |}
    \hline
    Ethereum Private Sidechain & \multicolumn{2}{c|}{Ethereum MainNet as Coordination Blockchain} \\
    Operation                            & \multicolumn{1}{|c|}{Advantages}      & \multicolumn{1}{|c|}{Disadvantages}   \\
    \hline
    \hline
    Discover using Ethereum    & Good authenticity, integrity, and                    &     \\
    Registration Authorities       & \hspace{3mm}non-repudiation properties. &  \\
                                                 & Permissionless, public network, &  \\
                                                 &  \hspace{3mm}enables discovery. &  \\
    \hline
    State Pinning \&                  &  Good authenticity, integrity, and                 &  Economic cost.      \\
    Final State Pinning              & \hspace{3mm}non-repudiation properties.  & Increased congestion on  \\
                                                &                                                                     &  \hspace{3mm}Ethereum MainNet. \\
                                                &                                                                      & Pinning and disputes are \\
                                                &                                                                      & \hspace{3mm}public.\\
    \hline
    State Pinning \&               & Leverage Ethereum MainNet                      &  Pins take more time to become  \\
    Final State Pinning          & \hspace{3mm}security properties while  & \hspace{3mm}\textit{final} on Ethereum MainNet   \\
    via an intermediate             & \hspace{3mm}minimising cost and congestion.     &  \hspace{3mm}than if pinned directly. \\
    private blockchain               & Pinning and disputes are not                      & Sidechain participants must  \\
                                                & \hspace{3mm}public.                                  & \hspace{3mm} observe all levels of pinning.\\
    \hline
    Sidechain Public Key          & Public keys widely available.   &  Significantly delays when first   \\
                                                &                      &  \hspace{3mm}Atomic Crosschain\\
                                                &                      &  \hspace{3mm}Transactions can be issued.\\
    \hline
    Atomic Crosschain              & Leverages Ethereum MainNet   &  Significantly delays when first   \\
    Transaction State                & \hspace{3mm}anti-spam capabilities.                &  \hspace{3mm}Atomic Crosschain\\
                                                &                                                          &  \hspace{3mm}Transactions can be issued.\\
                                                &                                                          &  Economic gas cost. \\
                                                &                                                            & Increased congestion on\\
                                                &                                                                     &  \hspace{3mm}Ethereum MainNet. \\
                                                
    \hline
    \hline
  \end{tabular}
  \caption{Advantages and Disadvantages of using Ethereum MainNet at Coordination Blockchain for Ethereum Private Sidechains}
  \label{table:merits}
\end{table*}

\subsection{Private Node Discovery - Ethereum Registration Authorities}
The Ethereum Registration Authorities system \cite{robinson2018b} uses smart contracts on a Coordination Blockchain to enable discovery of sidechain node address and cryptographic key information, as described in Section \ref{ref:ondemand}. As the information is used to bootstrap a sidechain, it is fundamental to the entire Ethereum Private Sidechain system that this information is authentic. 

The data in the Ethereum Registration Authority smart contracts is largely static. That is, the IP address and cryptographic key information, once set, changes rarely. Given this largely static data, the economic cost of storing information on Ethereum MainNet would only be incurred rarely. It is likely to cost less that US\$1.00 to set-up an enterprise in the Ethereum Registration Authority system on Ethereum MainNet, based on current prices  \cite{robinson2018b}.

Sidechain users who wish to establish a sidechain need to be able to access the bootstrap information stored in Ethereum Registration Authority smart contracts for the system to be useful. The information needs to be stored on a permissionless network or a permissioned network that has a black list of banned nodes. Doing this allows users who have no prior relationship with the operators of the Coordination Blockchain to access the information.

\subsection{State Pinning}
A private blockchain state pinning approach should be used to prevent state reversion as described in Section \ref{ref:ondemand}. Posting Pins to Ethereum MainNet leverages the authenticity, integrity and non-repudiation properties of Ethereum MainNet. However, submitting transactions costs money. Pinning once per hour for a year would cost US\$508 \cite{robinson2019a}. Additionally, if many sidechains pinned to Ethereum MainNet simultaneously, it would cause transaction congestion. Another issue with pinning directly to Ethereum MainNet is that any disputes that occur would need to occur on Ethereum MainNet, thus making the participant list of the sidechain public.

\begin{figure}
 \includegraphics[width=\linewidth]{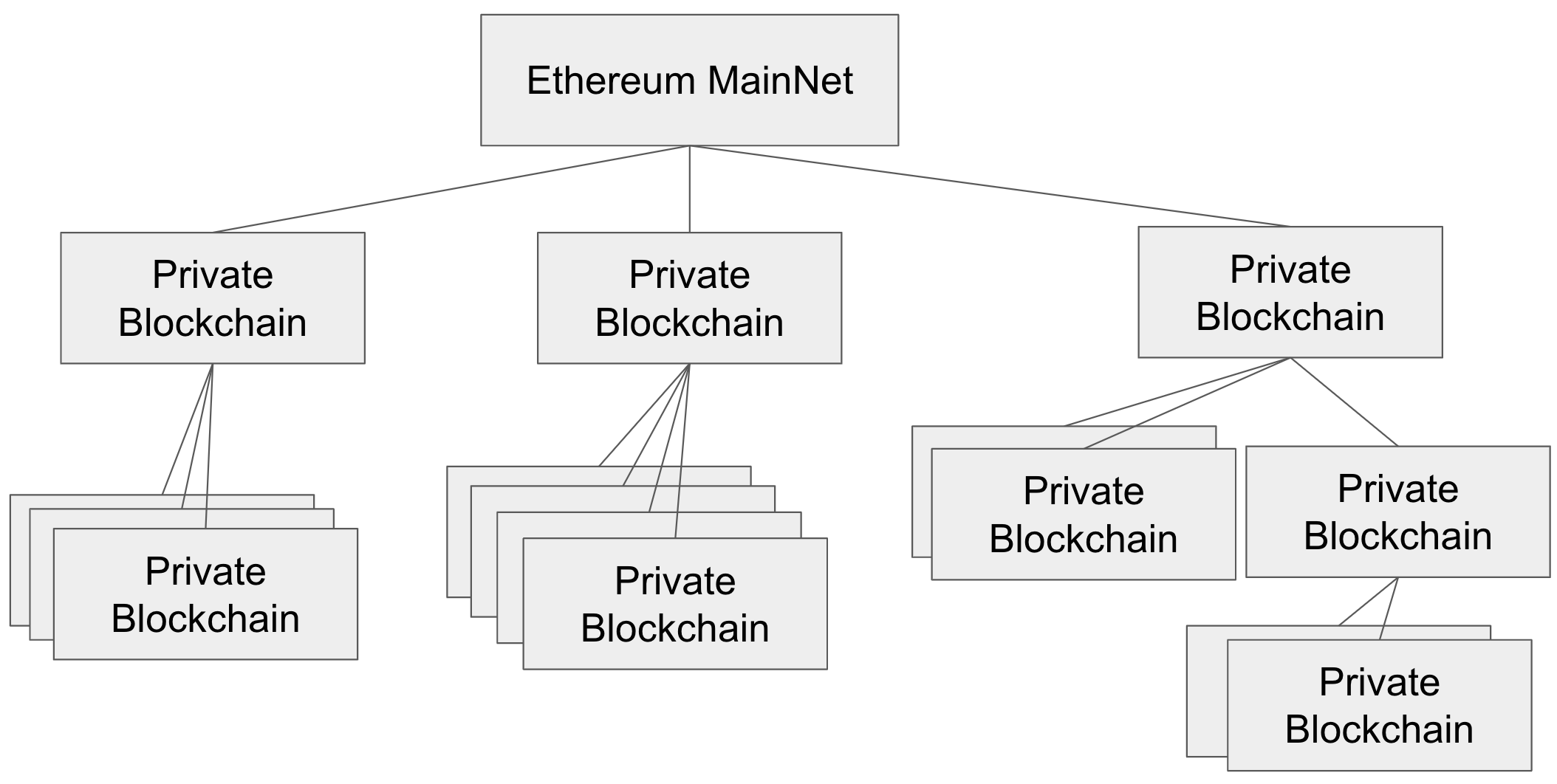}
 \caption{Hierarchical Pinning}
  \label{fig:hierarchical}
\end{figure}

Pins could be posted directly to a smart contract on Ethereum MainNet, or could be posted via a smart contract on an intermediate blockchain using a hierarchical pinning approach \cite{robinson2019a}. Using a hierarchical pinning approach, many private blockchains could treat another private blockchain as a Coordination Blockchain posting Pins to it. This private blockchain could in turn post Pins to another private blockchain or to Ethereum MainNet. This is shown diagrammatically in Figure \ref{fig:hierarchical}. Pinning to a hierarchy of Coordination Blockchains in this way means that only a small number of Pins on Ethereum MainNet could be used to secure a large number of private blockchains. The cost of submitting Pins to the private blockchain could be either free or significantly less than Ethereum MainNet. 

A benefit of pinning directly to Ethereum MainNet, rather than via an intermediate blockchain, is that the pinned state becomes \textit{final} faster. That is, if a Pin is posted to a private blockchain, whose state is in turn pinned to Ethereum MainNet, then the sidechain Pin could be deemed to become \textit{final} only once the private blockchain in pinned to Ethereum MainNet. 

Posting Pins via a private blockchain significantly reduces the cost of pinning, as only one blockchain needs to submit transactions to pin its state to Ethereum MainNet, and sidechains can pin to that private blockchain. Doing this reduces the number of transactions on Ethereum MainNet, thus reducing congestion, and means that the cost of submitting transactions is only incurred once for the private blockchain, rather than once for each sidechain.

A disadvantage of posting Pins via a private blockchain is that participants of the sidechain need to observe and be ready to challenge Pins being posted at each level of the hierarchy. If sidechain state Pins are posted directly to Ethereum MainNet, then the sidechain participants only need to observe the pinning contract on Ethereum MainNet.

An additional benefit of pinning to a private blockchain is that the chain's permissioning could be set such that only certain nodes could view the blockchain and only certain accounts could submit transactions to the blockchain. Pinning directly to Ethereum MainNet means that the organisation pinning to the contract is public. If there is a dispute, then masked participants will need to unmask themselves, and thus link themselves to the sidechain and the other organisations on the sidechain. If an intermediate blockchain was used, then the pinning and any disputes could happen in a more private setting.

\subsection{Final State Pinning for Archiving}
Final State Pinning is the same as State Pinning, with the exception that rather than the pinning being on an ongoing basis, it is just to pin the final state of a sidechain prior to archiving, as described in Section \ref{ref:ondemand}. As such, the advantages and disadvantages are similar to those described in the previous section. As only one pin is posted, the concerns over having to observe pins on a private blockchain in addition to Ethereum MainNet are not significant as the observation is for a single event. Similarly, concerns over cost of posting pins to Ethereum MainNet and congestion are reduced. As such, the advantages are reduced to the pin becoming final sooner and the disadvantages are reduced to any dispute over the value of the pin being public.

\subsection{Sidechain Public Keys}
\label{ref:sidechain-public-key}
As described in Section \ref{atomic}, the Atomic Crosschain Transactions feature needs Sidechain Public Keys to be stored on a Coordination Blockchain. The Sidechain Public Keys need to be stored in a contract \cite{robinson2019b} that allows voting on new public keys, and allows masked and unmasked participants. Given the participants are the same as those for the pinning scheme, it makes sense for these to be stored in the same contract as the pinning information. Keeping the logic in the same contract for pinning and holding the Sidechain Public Keys is useful as it means that membership changes need to only occur in one contract. However, the Sidechain Public Keys need to be visible by all sidechains that wish to verify information coming from the sidechain, whereas the pinning information need only be visible by sidechain participants and government regulators who would be appealed to in case of dispute.

Given the Sidechain Public Key is likely to be set once only, the economic cost of storing the key is likely to only be incurred once. No analysis of the gas cost of setting a Sidechain Public Key has been undertaken yet. However, given the small size of the public keys, \texttt{48} bytes, the incremental gas cost of storing the public key is likely to be in the order of \texttt{60,000} Gas, assuming the voting infrastructure has already been set-up. However, if the voting infrastructure did need to be set-up, the gas cost could be much larger. 

If a sidechain was short lived, then incurring the cost of setting up the voting infrastructure and posting the Sidechain Public Key to Ethereum MainNet could be deemed considerable. However, if the sidechain was long lived, then this relative cost might not be deemed as significant.

A disadvantage of using Ethereum MainNet to hold Sidechain Public Keys is transactions take at least \texttt{12} blocks before they should be deemed \textit{final} (see Section \ref{nonrep}). This means that, given a target block time of fourteen seconds, users could not use the Sidechain Public Keys for Atomic Crosschain Transactions for three minutes after the transaction that posts the Sidechain Public Key is included in a block on Etheurum MainNet.

\subsection{Atomic Crosschain Transaction State}
The Atomic Crosschain Transactions capability described in Section \ref{atomic} uses a Crosschain Coordination Contract to control when a crosschain transaction has started, been committed, or should be ignored. This information need to be available to all validators on all sidechains involved in the crosschain transaction. The information in the contract needs to be available until the last sidechain using the contract is archived. 

Storing the Atomic Crosschain State on Ethereum MainNet means that each Atomic Crosschain Transaction costs money to execute. This economic cost could be seen as an advantage, as it provides an anti-spam control external to the sidechain system. However, forcing enterprises to incur a cost for each crosschain transaction is likely to be viewed as an unnecessary cost. 

Additional issues with storing the Atomic Crosschain State on Ethereum MainNet is that this would leak the participants of a sidechain, as a transaction would need to be submitted linking the sidechain and the participant. Furthermore, this would leak the rate that the participant was issuing crosschain transactions. 

In a similar way that storing Sidechain Public Keys on Ethereum MainNet delays when the first Atomic Crosschain Transaction can be issued, as discussed in Section \ref{ref:sidechain-public-key}, storing Atomic Crosschain Transaction State could delay the effective start of each transaction. This is because sidechain participants might want to wait for blocks that contain transactions that indicate the Atomic Crosschain Transaction start to be final prior to acting on the start indication.

\section{Conclusion}
Coordination Blockchains perform various coordination tasks in private blockchain systems. We used Ethereum Private Sidechains as an exposition of such a system, highlighting the features of Ethereum Private Sidechains and discussing each feature's need to leverage a Coordination Blockchain. Based on the unique requirements of each feature and coordination activity, we examine whether public Ethereum MainNet would be a suitable platform for each of those tasks.

We found that Ethereum Registration Authority smart contracts of Ethereum Private Sidechains need to store long term data that have to be available in a permissionless blockchain. Ethereum MainNet would therefore be well suited to this task, as it is a permissionless blockchain that incentivises good behaviour using crypto economics, and provides good authenticity, integrity, and non-repudiation properties. Ethereum MainNet's strong security properties are also useful for State Pinning and in particular Final State Pinning, where the data needs to be stored securely for long periods of time. However, pinning directly to Ethereum MainNet could lead to congestion on Ethereum MainNet, would incur high costs, and would lead to the membership of a sidechain becoming public in the case of a dispute over the value of a Pin. These issues are significantly reduced by pinning via an intermediate private blockchain. However, doing this introduces other issues, such as participants having to observe pinned values at multiple levels in the pinning hierarchy and the pinned values taking longer to become final. Ethereum MainNet is not an appropriate location for Coordination Blockchain information that needs to be final quickly, such as Sidechain Public Keys and Atomic Crosschain Transaction State.

\ifarxiv
  \section*{Acknowledgments}
\fi
This research has been undertaken whilst I have been employed full-time at ConsenSys and have been completing my PhD part-time at University of Queensland. I acknowledge the support of my PhD supervisor Dr Marius Portmann. I thank Dr Catherine Jones, Horacio Mijail Anton Quiles, David Hyland-Wood, and Sandra Johnson for reviewing this paper and providing astute feedback.

\ifarxiv
\bibliographystyle{IEEEtran}
\bibliography{IEEEabrv,ref}
\fi

\end{document}